\documentclass[12pt,a4paper]{article}
\usepackage[cp866]{inputenc}
\usepackage{amsmath}
\usepackage{axodraw}
\usepackage{graphicx}
\newcommand{\be}{\begin{equation}}
\newcommand{\ee}{\end{equation}}
\newcommand{\ben}{\begin{equation*}}
\newcommand{\een}{\end{equation*}}
\newcommand{\bea}{\begin{eqnarray}}
\newcommand{\eea}{\end{eqnarray}}
\newcommand{\ar}{\begin{array}}
\newcommand{\arn}{\end{array}}

\newcommand{\vk}{\vec{k}}
\newcommand{\vks}{\vec{k}^{\;2}}
\newcommand{\q}{\vec{q}}
\newcommand{\qs}{\vec{q}^{\;2}}
\newcommand{\qp}{\vec{q}^{\;\prime}}
\newcommand{\qps}{\vec{q}^{\;\prime\; 2}}

\newcommand{\vl}{\vec{l}}

\def\pnot{\mbox{${\not{\hbox{\kern-3.0pt$p$}}}$}}
\def\qnot{\mbox{${\not{\hbox{\kern-2.0pt$q$}}}$}}
\def\enot{\mbox{${\not{\hbox{\kern-2.0pt$e$}}}$}}
\def\knot{\mbox{${\not{\hbox{\kern-2.0pt$k$}}}$}}

\def\fun#1#2{\lower3.6pt\vbox{\baselineskip0pt\lineskip.9pt\ialign
{$\mathsurround=0pt#1\hfil##\hfil$\crcr#2\crcr\sim\crcr}}}

\makeatletter 
\def\appendix{\par\clearpage 
  \setcounter{section}{0} 
  \setcounter{subsection}{0} 
  \@addtoreset{equation}{section} 
  \def\@sectname{Appendix~} 
  \def\theequation{\thesection.\arabic{equation}} 
  \def\thesection{\Alph{section}}} 
\makeatother 
 
\begin{document}
\sloppy                              
\renewcommand{\baselinestretch}{1.0} 

\begin{titlepage}
\hskip 11cm \vbox{ \hbox{Budker INP 2017-15}  }
\vskip 3cm

\begin{center}
{\bf Reggeon cuts in QCD amplitudes with negative signature$^{\ast}$}
\end{center}

\centerline{
V.S. Fadin $^{a,b}$, \fbox{L.N. Lipatov} $^{c,d}$}
\centerline{\sl $^{a}$
Budker Institute of Nuclear Physics of SD RAS, 630090 Novosibirsk
Russia}
\centerline{\sl\sl $^{b}$ Novosibirsk State University, 630090 Novosibirsk, Russia}
\centerline{
\sl $^{c}$Petersburg Nuclear Physics Institute, Gatchina, 188300, St.Petersburg, Russia}
\centerline{\sl $^{d}$ St.Petersburg State University, 199034, St.Petersburg, Russia}

\vskip 2cm

\begin{abstract}
Reggeon cuts  in QCD amplitudes with negative signature are discussed. These cuts appear   in the next-to-next-to-leading logarithmic approximation  and greatly complicate the derivation of the BFKL equation. Feynman diagrams responsible for appearance of these cuts are indicated  and the cut contributions are calculated in the two- and three-loop approximations.  
\end{abstract}


\vfill \hrule \vskip.3cm \noindent $^{\ast}${\it Work supported 
in part by the Ministry of Education and Science of Russian Federation,
in part by  RFBR,  grant 16-02-00888.}

\end{titlepage}

\section{Introduction}
One of the remarkable properties of  non-Abelian gauge theories (NAGTs) is  the  Reggeization of elementary particles in perturbation theory.  In contrast to Quantum Electrodynamics (QED), where the electron does Reggeize \cite{GGLMZ}, but the photon remains elementary \cite{M},  the criteria of Reggeization formulated  in \cite{M} are fulfilled in NAGTs for all particles \cite{Grisaru:1973vw, Grisaru:1974cf}.  In Ref.~\cite{Lipatov:1976zz},   by 
direct two-loop calculations in the leading logarithmic approximation (LLA), when in each order of perturbation theory only terms with the highest powers of the logarithm of c.m.s.  energy $\sqrt s$ are retained,  it was shown  that the gauge bosons of NAGTs do  Reggeize and   give the main contribution to the scattering amplitudes with the quantum numbers of  the gauge bosons and  negative signature  (symmetry with respect to the replacement $s\leftrightarrow u\simeq s$ )  in the $t$-channel. 
Then, based on  the three-loop calculations, it was assumed that the Regge form of such  amplitudes is valid in  the LLA in all orders of perturbation theory, and self-consistency of this assumption was checked \cite{Fadin:1975cb, Kuraev:1976ge}. A little bit later, the same was done for  the  amplitudes with the quark  quantum numbers  and  positive  signature in the $t$-channel \cite{Fadin:1976nw,  Fadin:1977jr}. Therefore, in Quantum Chromodynamics (QCD), which is a particular case of NAGTs, all elementary particles, i.e. quarks and gluons, do Reggeize.  

Reggeization of elementary particles is very important for the theoretical description of high energy processes.  The gluon  Reggeization is especially important because  it determines the high energy behaviour  of non-decreasing with energy cross sections in perturbative QCD. In particular, it appears to be the basis of the famous BFKL (Balitskii-Fadin-Kuraev-Lipatov) equation, which was first derived  in non-Abelian theories with spontaneously broken symmetry \cite{Fadin:1975cb, Kuraev:1976ge, Kuraev:1977fs}  and whose applicability in QCD was then shown  \cite{Balitsky:1978ic}.  In the BFKL approach the primary Reggeon is the Reggeized gluon. The Pomeron, which determines the high energy behaviour of cross sections, appears  as  a compound state of two Reggeized gluons, and  the Odderon, responsible for the difference of particle and antiparticle cross sections, as a compound state of three Reggeized gluons. 
 
The  Reggeization  allows  to express an infinite number of amplitudes   through several Reggeon vertices and the Reggeized gluon trajectory. 
It means definite  form not only
of elastic amplitudes, but  of inelastic amplitudes in the
multi-Regge kinematics  (MRK) as well.  This kinematics is very important because it gives  dominant contributions to cross sections and  to discontinuities of amplitudes with fixed momentum transfer in  the unitarity relations. In this kinematics  all particles have fixed (not growing with $s$) transverse momenta and are combined into jets with limited invariant mass of each jet and large (growing with $s$) invariant masses of any pair of the jets. In the LLA each jet contains only one gluon; in the next-to-leading logarithmic approximation (NLLA), where the eldest of non-leading terms are also retained,    one has to account production of $Q\bar Q$ and $GG$ jets.  

It is extremely important that in these approximations elastic amplitudes    and real parts of inelastic amplitudes in the MRK with negative  signature in cross-channels are determined by the Regge pole contributions   and have a simple factorized form (we will call it   pole Regge form).  Due to this, the Reggeization provides a simple derivation of the BFKL equation in the LLA and in  the NLLA. 

Validity of the Regge form  is proved now   in all orders of perturbation theory  in the coupling constant $g$ both  in the LLA \cite{Balitskii:1979}, and in the NLLA (see  \cite{Ioffe:2010zz, Fadin:2015zea} and references therein).

The  pole Regge form is violated in the NNLLA. The first observation of the violation was done  \cite{DelDuca:2001gu} at  consideration of the high-energy limit of the two-loop amplitudes for  $gg, gq$ and $qq$  scattering. The discrepancy appears in non-logarithmic   terms. If the pole Regge form would be correct in the NNLLA, they should satisfy  a definite condition (factorization condition), because
three amplitudes should be expressed in terms of two Reggeon-Particle-Particle vertices. 

Detailed consideration of  the terms responsible 
for   breaking of the pole Regge form in   two-loop and three-loop  amplitudes of elastic scattering  in QCD was performed in \cite{DelDuca:2013ara, DelDuca:2013dsa, DelDuca:2014cya}.  In particular, the  non-logarithmic terms violating the pole Regge form   at two-loops were  recovered  and not satisfying the factorization condition single-logarithmic terms  at  three loops were found using the  techniques of infrared factorization.  

It is necessary to say that, in general, breaking the pole Regge form is not a surprise.  It is well known that  Regge poles in the complex angular  momenta plane generate Regge cuts. Moreover, in amplitudes with positive signature the  Regge cuts appear already in the LLA. In particular, the BFKL  Pomeron  is the two-Reggeon cut in  the complex angular  momenta plane.  But in amplitudes with  negative signature Regge cuts appear only in the NNLLA.  It is   natural to expect that the observed violation of the  pole Regge form can be  explained by their contributions.   As it is  shown in  \cite{Fadin:2016wso, Caron-Huot:2017fxr}, this is actually so.  

Here  the results on which the report \cite{Fadin:2016wso}  at  the workshop "Diffraction 2016" was based are presented.  Since for  violation of the  pole Regge  form  only  amplitudes with the gluon quantum numbers in the $t$-channel and  negative signature are  important, only  such amplitudes are  considered below.  The three-Reggeon cuts in other channels with  negative signature are discussed in \cite{Caron-Huot:2017fxr, Fadin:2017}

\section{Lowest order contribution}  
\begin{figure}[h]
\begin{minipage}[l]{50mm}
\begin{center}
\begin{picture}(100,100)(0,0)
\ArrowLine(10,90)(50,90) 
\ArrowLine(50,90)(90,90) \Text(10,100)[c]{$A$}
\Text(90,100)[c]{${A'}$} 
\Gluon(35,90)(35,10){3}{11}
\Gluon(50,90)(50,10){3}{11}
\Gluon(65,90)(65,10){3}{11}
\ArrowLine(10,10)(50,10) 
\ArrowLine(50,10)(90,10)
\Text(10,0)[]{$B$} \Text(90,0)[]{${B'}$} \Text(50,-10)[]{$a$} 
\end{picture}
\end{center}
\end{minipage}
\begin{minipage}[l]{50mm}
\begin{center}
\begin{picture}(100,100)(0,0)
\ArrowLine(10,90)(50,90) 
\ArrowLine(50,90)(90,90) \Text(10,100)[c]{$A$}
\Text(90,100)[c]{${A'}$} 
\Gluon(35,90)(35,10){3}{11}
\Gluon(50,90)(65,10){3}{11}
\Gluon(65,90)(50,10){3}{11}
\ArrowLine(10,10)(50,10) 
\ArrowLine(50,10)(90,10)
\Text(10,0)[]{$B$} \Text(90,0)[]{${B'}$} \Text(50,-10)[]{$b$} 
\end{picture}
\end{center}
\end{minipage}
\begin{minipage}[l]{50mm}
\begin{center}
\begin{picture}(100,100)(0,0)
\ArrowLine(10,90)(50,90) 
\ArrowLine(50,90)(90,90) \Text(10,100)[c]{$A$}
\Text(90,100)[c]{${A'}$} 
\Gluon(35,90)(50,10){3}{11}
\Gluon(50,90)(35,10){3}{11}
\Gluon(65,90)(65,10){3}{11}
\ArrowLine(10,10)(50,10) 
\ArrowLine(50,10)(90,10)
\Text(10,0)[]{$B$} \Text(90,0)[]{${B'}$} \Text(50,-10)[]{$c$} 
\end{picture}
\end{center}
\end{minipage}

\vspace{10mm}

\begin{minipage}[l]{50mm}
\begin{center}
\begin{picture}(100,100)(0,0)
\ArrowLine(10,90)(50,90) 
\ArrowLine(50,90)(90,90) \Text(10,100)[c]{$A$}
\Text(90,100)[c]{${A'}$} 
\Gluon(35,90)(50,10){3}{11}
\Gluon(50,90)(65,10){3}{11}
\Gluon(65,90)(35,10){3}{11}
\ArrowLine(10,10)(50,10) 
\ArrowLine(50,10)(90,10)
\Text(10,0)[]{$B$} \Text(90,0)[]{${B'}$} \Text(50,-10)[]{$d$} 
\end{picture}
\end{center}
\end{minipage}
\begin{minipage}[l]{50mm}
\begin{center}
\begin{picture}(100,100)(0,0)
\ArrowLine(10,90)(50,90) 
\ArrowLine(50,90)(90,90) \Text(10,100)[c]{$A$}
\Text(90,100)[c]{${A'}$} 
\Gluon(35,90)(65,10){3}{11}
\Gluon(50,90)(35,10){3}{11}
\Gluon(65,90)(50,10){3}{11}
\ArrowLine(10,10)(50,10) 
\ArrowLine(50,10)(90,10)
\Text(10,0)[]{$B$} \Text(90,0)[]{${B'}$} \Text(50,-10)[]{$e$} 
\end{picture}
\end{center}
\end{minipage}
\begin{minipage}[l]{50mm}
\begin{center}
\begin{picture}(100,100)(0,0)
\ArrowLine(10,90)(50,90) 
\ArrowLine(50,90)(90,90) \Text(10,100)[c]{$A$}
\Text(90,100)[c]{${A'}$} 
\Gluon(35,90)(65,10){3}{11}
\Gluon(50,90)(50,10){3}{11}
\Gluon(65,90)(35,10){3}{11}
\ArrowLine(10,10)(50,10) 
\ArrowLine(50,10)(90,10)
\Text(10,0)[]{$B$} \Text(90,0)[]{${B'}$} \Text(50,-10)[]{$f$} 
\end{picture}
\end{center}
\end{minipage}

\vspace{5mm}

\caption{Three-gluon exchange diagrams} \label{Fig:3g}

\end{figure}
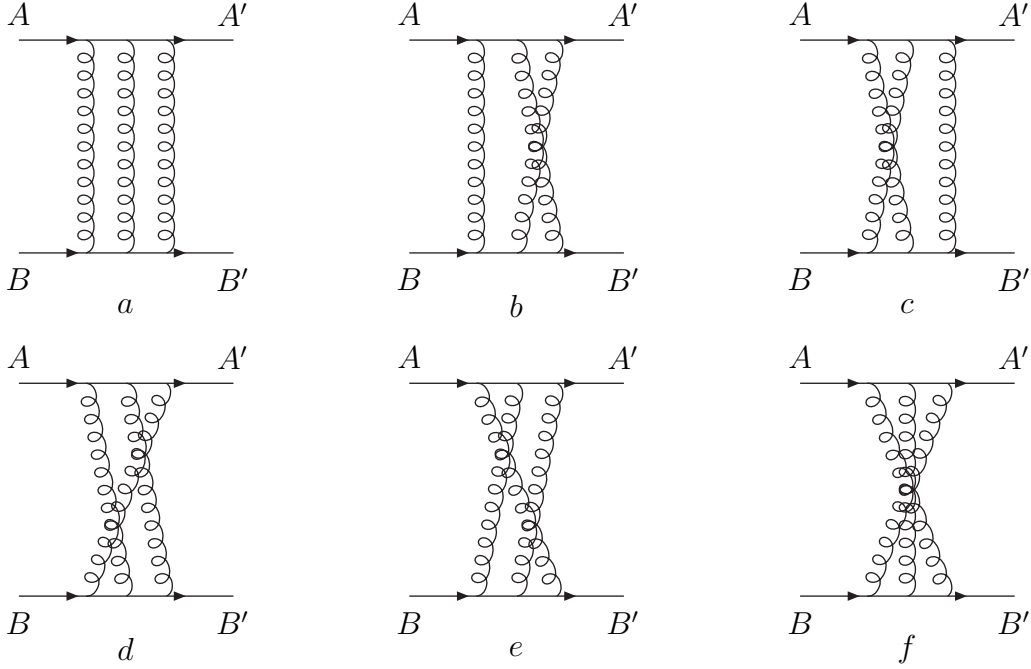

Let us consider parton (quark and gluon) elastic scattering amplitudes with   negative signature in the two-loop approximation, and try to find the contribution  of the Regge cut in them. Due to the signature conservation the cut with  negative signature  has to be a three-Reggeon one.  Since our Reggeon is the Reggeized gluon, the cut starts with the diagrams with three $t$-channels gluons.  They are presented in Fig.\ref{Fig:3g}, where particles $A, A'$ and $B, B'$ can be quarks or gluons.

\noindent 
The colour structures of the diagrams in Fig.\ref{Fig:3g} can be decomposed into irreducible  representations of the colour group in the $t$-channel. Since  we confine ourselves to  amplitudes with  the gluon quantum numbers in the $t$-channel,  we  need  to  consider only the  adjoint representations.  Moreover, since we are interested in the  negative signature,  for gluon scattering we need  to consider only   the  antisymmetric adjoint representation.   In other words, in the colour decomposition  we  need only  the same  structure as in the one-gluon exchange,  
\be
C_{AB}^{A'B'} =\langle A'|{\cal T}^a|A\rangle \langle B'|{\cal T}^a|B\rangle~, \label{born colour}
\ee 
where ${\cal T}^a$ are  the colour group generators in the corresponding representations, $[{\cal T}^a, {\cal T}^b] =if_{abc}  {\cal T}^c$;   ${\cal T}^a_{bc} = T^a_{bc} = -if_{abc} $ for gluons and ${\cal T}^a_{bc} = -t^a_{bc}$ for quarks. 

Matrix elements corresponding to the diagrams in Fig.\ref{Fig:3g} contain this colour structure with the  coefficients 
$C^\alpha_{ij}$, where $\alpha =a, b, c,d,e,f$ and $ij=gg, gq$ and $qq$
for gluon-gluon, gluon-quark and quark-quark scattering correspondingly. They are given  by the convolutions 
$\mbox{Tr}({\cal T}^a{\cal T}^b{\cal T}^c{\cal T}^d)\mbox{Tr}({\cal T}^{a_1}{\cal T}^{b_1}{\cal T}^{c_1}{\cal T}^d)$, where $a_1b_1c_1$ are obtained from $abc$ by all possible permutations.  

\begin{figure}[h]
\begin{center}
\begin{picture}(160,160)(-10,-10)
\ArrowLine(75,160)(75,-10) 
\Text(70,155)[c]{$q$}
\ArrowArc(225, 75)(165, 156, 204)
\ArrowArcn(-75, 75)(165, 24, -24)
\Text(70,0)[c]{$q$}
\end{picture}
\end{center}

\vspace{5mm}
\caption{Diagrammatic representation of $A^{(3)}_\perp$ in the transverse momentum space }\label{Fig:3 transverse}

\end{figure}
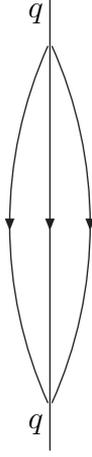

Using the equalities 
\be
\mbox{ Tr}(t^a t^b) = \frac12 \delta^{ab}~, \;\; 
t^a t^b  =\frac{1}{2N_c}\delta^{ab} I + \frac12 (d^{abc}+if^{abc})t^c~, \;\; \label{tatb}
\ee
one easily finds 
\be
\mbox{Tr}(t^at^bt^ct^d)= \frac{1}{{N_c}}\delta^{ad}\delta^{bc} +\frac{1}{8}(d^{adi}d^{bci}+f^{adi}f^{bci} +id^{adi}f^{bci} -if^{adi}d^{bci})~. \label{tttt} 
\ee
As it follows from \eqref{tatb}, the tensor $\mbox{Tr}(T^aT^bT^cT^d)$ can be written as 
\be
\mbox{Tr}(T^aT^bT^cT^d) = 4 \mbox{Tr}([t^a,t^i][t^j, t^b])\mbox{Tr}([t^c,t^j][t^i, t^d])~. \label{TaTbTcTd}
\ee
Using this representation, the completeness condition for the matrices $t^a$ and the identity matrix $I$ in the form 
\be
 \mbox{Tr}(t^i A) \mbox{Tr}(t^i B) = -\frac{1}{2N_c}\mbox{Tr}(A) \mbox{Tr}(B) +\frac{1}{2}\mbox{Tr}(AB)~, \label{completeness}
\ee 
and the  relations (which are easily obtained from the completeness condition) 
\be
t^at^a = C_F I, \;\;  t^at^bt^a
=\left(C_F-\frac{C_A}{2}\right)t^b,\;\;t^at^bt^ct^a
=\frac{1}{4}\delta^{bc}I +\left(C_F-\frac{C_A}{2}\right)t^bt^c~, \label{ttt}
\ee
where $C_F$ and $C_A$ are the values of the Casimir operators in
the fundamental and adjoint representations, 
\be\label{[CF CA]}
C_F=\frac{N_c^2-1}{2N_c}\,, \qquad C_A=N_c\,, 
\ee
we obtain
\be
\mbox{Tr}(T^aT^bT^cT^d)= \delta^{ad}\delta^{bc}+\frac{1}{2}(\delta^{ab}\delta^{cd}+\delta^{ac}\delta^{bd}) +\frac{{N_c}}{4}(f^{adi}f^{bci}+d^{adi}d^{bci})~.  \label{TaTbTcTd-1}
\ee
The  convolutions can be performed with the help of the relations 
\be
\mbox{Tr}\left(T^aD^b\right)= 0~, \;\;  \;\;\mbox{Tr}\left(T^aT^b\right)=N_c\delta^{ab},\ \quad
\mbox{Tr}\left(D^aD^b\right)=\frac{N^2_c-4}{N_c}\delta^{ab}\,,
\ee
\be
\mbox{Tr}\left(T^aT^bT^c\right)=i\frac{N_c}{2}f^{abc},\ \quad
\mbox{Tr}\left(T^aT^bD^c\right)=\frac{N_c}{2}d^{abc}\,,
\ee
\be
\mbox{Tr}\left(D^aD^bT^c\right)=i\frac{N^2_c-4}{2N_c}f^{abc},\quad
\mbox{Tr}\left(D^aD^bD^c\right)=\frac{N^2_c-12}{2N_c}d^{abc},
\ee
where $D^a_{bc}= d_{abc}$. They can be derived analogously to \eqref{TaTbTcTd}. 
As the result, one obtains
\be
C^a_{gg} =\frac32+\frac{N_c^2}{8}~, \; C^a_{gq} =\frac14+\frac{N_c^2}{8}~, \;  C^a_{qq} =\frac14\left(1+\frac{3}{N_c^2}\right)~, 
\ee
\be
C^b_{gg} =C^c_{gg}=C^d_{gg} =C^e_{gg} =C_{gg} =\frac32~,\;  C^b_{gq} =C^c_{gq}=C^d_{gq} =C^e_{gq}=  C_{gq} =\frac14~, 
\ee
\be
 C^b_{qq} =C^c_{qq}=C^d_{qq} =C^e_{qq}  = C_{qq} =\frac14\left(-1+\frac{3}{N_c^2}\right)~, 
\ee
\be
C^f_{gg} =\frac32+\frac{N_c^2}{8}~,\; C^f_{gq} =\frac14+\frac{N_c^2}{8}~, \;  C^f_{qq} =\frac14\left(N_c^2-3+\frac{3}{N_c^2}\right)~,    \label{coefficients Cij}
\ee
The contribution $A^{Fig.1}$ of the diagrams in Fig.\ref{Fig:3g} to the  scattering amplitudes with  the colour structures \eqref{born colour} can be written  as 
\be
A_{ij}^{Fig.1} =\langle A'|{\cal T}^a|A\rangle \langle B'|{\cal T}^a|B\rangle\left[ C_{ij} A^{eik} + \frac{N_c^2}{8} \left(A_{ij}^{a}+A_{ij}^{f}\right)+ \delta_{i,q}\delta_{j,q}\frac{4-N_c^2}{8} \left(A_{ij}^{a}-A_{ij}^{f}\right)\right]~, \label{Aij}
\ee   
where $A_{ij}^{\alpha}$ is the contribution of the diagram $\alpha$ with omitted colour 
factors and  $A_{ij}^{eik} = \sum_\alpha A_{ij}^{\alpha}$. Note that $A^{eik}$ is gauge invariant.  It can be easily found: 
\be 
A^{eik} = g^2\frac{s}{t}\left(\frac{-4\pi^2}{3}\right)g^4\,\qs\, A^{(3)}_\perp ~, 
\ee
where $A^{(3)}_\perp $ is presented by the diagram in Fig.\ref{Fig:3 transverse} in the transverse momentum space, and is given by the integral  
\[
A^{(3)}_\perp
 = \int\frac{{d^{2+2\epsilon}l_1}{d^{2+2\epsilon} l_2}}{(2\pi)^{2(3+2\epsilon)}\vec l_1^{\:
2}\vec l_2^{\:
2}(\vec q -\vec l_1-\vec l_2)^2}=
\]
\be
= 3C_{\Gamma}^2\frac{4}{\epsilon^2}\frac{(\vec q^{\:
2})^{2\epsilon}}{\vec q^{\:
2}} \frac{\Gamma^2(1+2\epsilon)\Gamma(1-2\epsilon)}
{\Gamma(1+\epsilon)\Gamma^2(1-\epsilon)\Gamma(1+3\epsilon)}~,
\ee
where 
\be
C_{\Gamma}= \frac{\Gamma(1-\epsilon)\Gamma^2(1+\epsilon)}
{(4\pi)^{2+\epsilon}\Gamma(1+2\epsilon)} = \frac{\Gamma(1-\epsilon)\Gamma^2(1+\epsilon)}
{(4\pi)^{2+\epsilon}}(1-\epsilon^2 \zeta(2)+2\epsilon^3 \zeta(3)-\frac94\epsilon^4 \zeta(4)+.....)~,\;\;\;
\ee 
\be
\frac{\Gamma^2(1+2\epsilon)\Gamma(1-2\epsilon)}
{\Gamma(1+\epsilon)\Gamma^2(1-\epsilon)\Gamma(1+3\epsilon)}=1+ 6\epsilon^3 \zeta(3)-9\epsilon^4 \zeta(4)+.....)~.
\ee
Note that we use the "infrared" $\epsilon$, $\epsilon= (D-4)/2$, $D$ is the space-time dimension. 

The last term in \eqref{Aij} is not relevant; it is  the contribution of  positive signature in the quark-quark scattering.   The second term does not violate the pole factorization and can be assigned to the Reggeized gluon contribution. This is not true
for  the first term, because 
\be
2C_{gq} \neq  C_{qq}+C_{gg}~, \;\; 2C_{gq} - C_{qq} - C_{gg} = -\frac14\left(1+\frac{1}{N_c^2}\right)~, 
\ee
which means violation of  the pole factorization. It is not difficult to see  that the nonvanishing in the limit $\epsilon \rightarrow 0$ part of the amplitudes $C_{ij}A^{(eik)}$ coincides with  $g^2(s/t)(\alpha_s/\pi)^2 {\cal R}_ij^{(2), 0, [8]}$ of the paper \cite{DelDuca:2014cya}.  The values  ${\cal R}_ij^{(2), 0, [8]}$ are given there in  Eq. (4.35),  $(\alpha_s/\pi)^2 {\cal R}_ij^{(2), 0, [8]}$   is the first (two-loop, non-logarithmic) contributions to the "non-factorizing   remainder function" ${\cal R}^{[8]}_{ij}$ introduced in  Eq. (3.1)  of \cite{DelDuca:2014cya}.  This means that the violation of the  pole factorization discovered  in \cite{DelDuca:2001gu} and analysed in  \cite{DelDuca:2013ara}-\cite{DelDuca:2014cya} is due  to the eikonal part of the contribution of the diagrams with three-gluon exchange. 

However, one can not affirm that this part  is given entire by the three-Reggeon cut. 
Indeed, it can contain also  the Reggeized gluon contribution. In fact,  a  non-factorizing   remainder function is not  uniquely defined. The definition used in \cite{DelDuca:2014cya}  was chosen for convenience of comparison of the high energy and infrared factorizations. 
\section{Radiative corrections}
\begin{figure}[h]
\begin{center}
\begin{picture}(160,160)(-10,-10)
\ArrowLine(75,160)(75,140) 
\Text(70,155)[c]{$q$}
\ArrowArc(225, 75)(165, 156, 204)\ArrowArc(550, 75)(480, 171, 189)
\ArrowArcn(-75, 75)(165, 24, -24)\ArrowArcn(-400, 75)(480, 9, -9)
\ArrowLine(75,10)(75,-10) 
\Text(70,0)[c]{$q$}
\ArrowLine(75,10)(75,-10)
\end{picture}
\end{center}

\vspace{5mm}
\caption{Diagrammatic representation of $A^{(3)}_\perp C_R$ }\label{Fig:4 transverse}
\end{figure}
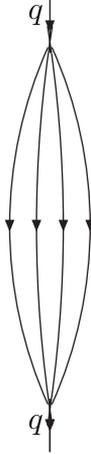

The problem of 
separation of the  pole and cut contributions can be solved by consideration of 
logarithmic radiative corrections to them. In the case of the Reggeized gluon contribution the correction comes  solely from the Regge factor, so that  the first order correction (more strictly, its relative value; this is assumed also in the following) is $\omega(t)\ln s$, where $\omega(t)$ is the gluon trajectory, 
\be
\omega(t)
 =-g^2 N_c {\vec q^{\:
  2}} \int\frac{d^{2+2\epsilon}l}{2(2\pi)^{(3+2\epsilon)}\vec l^{\:
2}(\vec q -\vec l)^2}= -g^2 N_c C_{\Gamma}\frac{2}{\epsilon}(\vec q^{\:
2})^{\epsilon} ~. \label{omega}
\ee
In the case of the three-Reggeon cut, one has to take into account the Reggeization of each  of   three gluons and the interaction between them. The Reggeization 
gives $\ln s$ with the coefficient  $3C_R$, where 
$A^{(3)}_\perp C_R$ is represented by the diagram in Fig.\ref{Fig:4 transverse} in the transverse momentum space,

and is given by the integral  
\[
A^{(3)}_\perp C_R =  -g^2 N_c C_{\Gamma}\frac{2}{\epsilon}  \int\frac{d^{2+2\epsilon}l_1\,d^{2+2\epsilon} l_2}{(2\pi)^{2(3+2\epsilon)}\vec l_1^{\:
2}\vec l_2^{\:
2}(\vec q -\vec l_1-\vec l_2)^{1-\epsilon}}
\]
\be
=  -g^2 N_c C_{\Gamma}\frac{4}{3\epsilon}  (\vec q^{\:
2})^{\epsilon}
\frac{\Gamma(1-3\epsilon)\Gamma(1+2\epsilon)\Gamma(1+3\epsilon)}
{\Gamma(1-\epsilon)\Gamma(1-2\epsilon)\Gamma(1+\epsilon)\Gamma(1+4\epsilon)}A^{(3)}_\perp~. \label{CR}
\ee
Interaction between two  Reggeons with transverse momenta $\vl_1$ and $\vl_2$  and colour indices $c_1$ and $c_1$ is defined  by the real part of the BFKL kernel 
\be
\left[{\cal K}_r(\q_1, \q_{2}; \vk)\right]^{c'_1c'_2}_{c_1c_2} =
T_{c_1c'_1}^{a}T_{c_2c'_2}^{a}\frac{g^2}{(2\pi)^{D-1}}
\left[\frac{\qs_{1}\qps_{2}+\qs_{2}\qps_{1}}
{\vks}-\qs\right], \label{explicit K r}
\ee
where  $\vk$ is the  momentum transferred  from one Reggeon to another in the  interaction, $\qp_1$  and $\qp_2$ ($c'_1$ and $c'_2$) are the Reggeon momenta (colour indices) after the interaction, $\qp_1=\q_1 -\vk, \;\; \qp_2=\q_2 +\vk, $  and $\q =\q_1+\q_2=\qp_1+\qp_2$. 
  
For the colour structure which we are interested in,   account of the interactions between all  pairs  of  Reggeons  leads in the sum to the colour coefficients which  differ from the coefficients $C^{\alpha}_{ij}$ \eqref{coefficients Cij}
only by the common factor $N_c$. It  can be easily obtained using  invariance  of $\mbox{Tr}( {\cal T}^{a_1}{\cal T}^{a_2}{\cal T}^{a_3}{\cal T}^{a_4})$  under the  colour group transformation
\be
{\cal T}^{a_i}\rightarrow e^{i\theta^c {\cal T}^{c} }
{\cal T}^{a_i}e^{-i\theta^c {\cal T}^{c} }~, 
\ee
which takes the form 
\be
{\cal T}^{a_i}\rightarrow {\cal T}^{a_i} -i \theta^c \hat T^c(i){\cal T}^{a_i}~, \;\;  
\hat T^c(i){\cal T}^{a_i} = T^c_{a_ia'_i}{\cal T}^{a'_i} 
\ee
at small  $\theta^c $.   It means that we can put 
\be
\hat R^c = \sum_i \hat T^c (i)  =0~,  \label{Rc=0}
\ee
if $\hat R^c$ acts on   $\mbox{Tr}( {\cal T}^{a_1}{\cal T}^{a_2}{\cal T}^{a_3}{\cal T}^{a_4})$. Using \eqref{Rc=0} and $\hat R^c \hat R^c =0 $ one has 
\be
\sum_{i>j =2}^4 \hat T^c (i) \hat T^c (j) =  \frac12 \left(\sum_{i=2}^4 \hat T^c (i) \hat T^c (i)
- \hat T^c (1) \hat T^c (1)\right)~.  
\ee
Here on the left side we have the sum of the colours factors  of  the BFKL kernel \eqref{explicit K r} for interactions  between all  pairs  of  Reggeons, and the  right  side is equal $N_c$. 

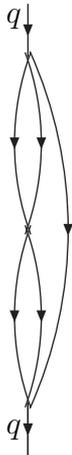
\begin{figure}[h]
\begin{center}
\begin{picture}(160,160)(-10,-10)
\ArrowLine(75,160)(75,140) 
\Text(70,155)[c]{$q$}
\ArrowArcn(-75, 75)(165, 24, -24)

\ArrowArcn(-20, 42.5)(100, 20, -20)\ArrowArc(170, 42.5)(100, 160, 200) 
\ArrowArcn(-20, 107.5)(100, 20, -20)\ArrowArc(170, 107.5)(100, 160, 200)

\ArrowLine(75,10)(75,-10) 
\Text(70,0)[c]{$q$}
\ArrowLine(75,10)(75,-10)

\end{picture}
\end{center}

\vspace{5mm}
\caption{Diagrammatic representation of $A^{(3)}_\perp C_R$ }\label{Fig:5 transverse}
\end{figure}

Now about the kinematic part of the kernel \eqref{explicit K r}. The first two terms in the square brackets in  \eqref{explicit K r} correspond  to the  same diagrams  
as in \eqref{Fig:4 transverse}, and their total contribution  to the coefficient of 
$\ln s$  in the first order correction is $-4C_R$. The last term  corresponds to the 
diagram in Fig.\ref{Fig:5 transverse} 

\vspace{5mm}

and its contribution to the coefficient of  $\ln s$   in the first order correction is $-C_3$, where 
\[
C_3 =  g^2 N_c C_{\Gamma}\frac{4}{\epsilon}  \int\frac{d^{2+2\epsilon}l_1\,d^{2+2\epsilon} l_2 (\l_1+\l_2)^{2\epsilon}}{(2\pi)^{2(3+2\epsilon)}\vec l_1^{\:
2}\vec l_2^{\:
2}(\vec q -\vec l_1-\vec l_2)^{1-2\epsilon}}\left(A^{(3)}_\perp\right)^{-1}
\]
\be
=  g^2 N_c C_{\Gamma}\frac{32}{9\epsilon}  (\vec q^{\:
2})^{\epsilon}
\frac{\Gamma(1-3\epsilon)\Gamma(1-\epsilon)\Gamma^2(1+3\epsilon)}
{\Gamma^2(1-2\epsilon)\Gamma(1+2\epsilon)\Gamma(1+4\epsilon)}~.\label{C3}
\ee
Therefore, the first order correction in the case of Reggeized gluon is $\omega(t)\ln s$, where $\omega(t)$ is given by \eqref{omega},
and in the case of the three-Reggeon  cut  is $(-C_R -C_3)\ln s$, where $C_R$ and $C_3$ are given by \eqref{CR} and  \eqref{C3} respectively.  If to  present the  coefficients $C_{ij}$ in  \eqref{coefficients Cij}  as the sum 
\be
C_{ij} = C^R_{ij}+C^C_{ij}~, \label{C=CR+CC} 
\ee
where $C^R_{ij}$ correspond to the pole, so that 
\be
2C^R_{gq} =   C^R_{qq}+C^R_{gg}~, \label{2C=C+C} 
\ee
and $C^C_{ij}$ correspond to the cut, we obtain that  with the logarithmic accuracy  the total three-loop contributions to the coefficient of  $\ln s$ are
\be
A^{eik}\left(C^R_{ij} \omega(t)-C^C_{ij}(C_R+C_3)\right)\ln s~. \label{total}
\ee
The infrared divergent part of these   contributions must be compared with the functions  $g^2(s/t){\cal R}_{ij}^{(3), 1, [8]} \ln s$ of the paper \cite{DelDuca:2014cya}.  The values  ${\cal R}_ij^{(3), 1, [8]}$ are given there in  Eq. (4.59), ${\cal R}_ij^{(3), 1, [8]}\ln s$ are  the three-loop logarithmic contributions to the non-factorizing   remainder function ${\cal R}^{[8]}_{ij}$.  
It is not difficult to see  that with the accuracy with which the values ${\cal R}_ij^{(2), 1, [8]}$ are known the equality 
\be 
g^2(s/t){\cal R}_{ij}^{(3), 1, [8]} = A^{eik}\left(C^R_{ij} \omega(t)-C^C_{ij}(C_R+C_3)\right)
\ee
can be fulfilled if 
\be
 C^C_{gg} = -\frac32~,  \; \;C^C_{gq} =-\frac{3}{2}~, \;  \; \;C^C_{qq} =\frac{3(1-N_c^2)}{4N_c^2}~,   \label{coefficients CCij}
\ee
and
\be 
C^R_{gg} = 3~, \; \; C^R_{gq} =\frac{7}{4}~, \; \; C^R_{qq} =\frac{1}{2}~.\label{coefficients CRij}
\ee
It means that the restrictions imposed by the infrared factorization  on the parton scattering amplitudes with the adjoint representation of the colour group in the $t$-channel and  negative signature   can be fulfilled in the NNLLA at two  and three loops if besides the Regge pole contribution there is the Regge cut contribution 
\be
A^{eik}C^C_{ij}\left(1-(C_R+C_3)\ln s\right)~. \label{full}
\ee
Here the coefficients  $C^C_{ij}, \; C_R$ and  $C_3$ are given by Eqs. \eqref{coefficients CCij},  \eqref{CR} and  \eqref{C3} respectively. 

\section{On the derivation of the BFKL equation in the NNLLA}
The BFKL equation was derived \cite{Fadin:1975cb, Kuraev:1976ge}  for summation of radiative corrections in the LLA to  amplitudes of elastic scattering processes.  These amplitudes  were calculated using   the  $s$-channel unitarity and analyticity. The unitarity was used for calculation of discontinuities  of  elastic amplitudes, and analyticity  for their full restoration.  Use of the  $s$-channel unitarity  requires knowledge of multiple production amplitudes in the MRK.  The assumption was made that all amplitudes in the unitarity relations for elastic amplitudes, both elastic and inelastic, are determined  by the Regge pole contributions.    With this assumption, the $s$-channel discontinuities  of the elastic amplitudes  can be presented as the convolution in the transverse momentum space  of energy independent  impact factors of  colliding particles,  describing their interaction   with Reggeons,   and  the  Green's function $G$ for two interacting Reggeons, which is universal (process independent). The BFKL equation looks as 
\be
\frac{d\;G}{d\ln s}  = \hat{\cal{K}} \; G~, 
\ee
where  ${\hat{\cal{K}}}$ is the  kernel of the BFKL equation.    It consists of  virtual and real parts; the first of them 
is expressed through the Regge trajectories and the second through effective vertices  for    $s$-channel production of particles in Reggeon  interaction. 

The assumption  that all amplitudes in the unitarity relations are determined  by the Regge pole contributions is very strong, and it should have been proven. The first check of this assumption was made already in \cite{Kuraev:1976ge}. Here it should be recalled that  the BFKL equation is written for  all $t$-channel colour states, which  a system of two Reggeons can have, in particular, for the colour octet.  Therefore, there is the bootstrap requirement: 
solution of the BFKL equation for the colour octet in the $t$-channel and negative signature  must reproduce the  pole Regge form, which was assumed in its  derivation. It was shown in \cite{Kuraev:1976ge} that this requirement is satisfied. Of course, it was not a proof of the assumption, but only  verification  in a very particular case. Later it was realized that it is possible to formulate the bootstrap conditions for amplitudes of multiple production in the MRK and to give a complete proof of the hypothesis   on their basis  \cite{Balitskii:1979}. A similar (although much more complicated) proof for elastic amplitudes and for real parts of inelastic ones was carried out in  the NLLA (see \cite{Ioffe:2010zz, Fadin:2015zea} and references therein).

It turns out that all amplitudes in the unitarity relations  are determined  by the Regge pole contributions also in  the  NNLLA. The reason is that in this approximation one of two amplitudes in the unitarity relations  can lose $\ln s$, while the second one must be taken in the LLA.  The LLA amplitudes are real, so that only real parts of the NLLA amplitudes are important in the unitarity relations. Since they have  a simple  pole Regge form, the scheme of deviation of the BFKL equation in the NLLA remains unchanged. The only difference is that we have to know the Reggeon trajectory and Reggeon-Reggeon-gluon production vertex with higher accuracy and to know also effective Reggeon-Reggeon$\rightarrow$ gluon-gluon and  Reggeon-Reggeon$\rightarrow$ quark-antiquark vertices.

Unfortunately, this scheme  is violated in  the NNLLA.  In this approximation two powers of $\ln s$ can be lost compared with the LLA in the product of two amplitudes in the unitarity relations. It can be done  losing either one $\ln s$ in each of the amplitudes or $\ln^2 s$ in one of them.   In the first case, discontinuities receive contributions from  products of real parts of amplitudes with negative signature in the NLLA, products of imaginary parts of  amplitudes with negative signature in the LLA, and products of amplitudes  with positive signature in the LLA. Of course,  account of these contributions greatly complicates derivation of the BFKL equation. In particular, since for amplitudes with positive signature there are different colour group representations  in the $t$-channel for quark-quark, quark-gluon and gluon-gluon scattering, their account violates  unity  of  consideration.

However, these complications do not seem to be as great as in the second case, when  $\ln^2 s$ is lost in one of the amplitudes in the unitarity relations. In this  case  one of the amplitudes must be taken in the NNLLA and the other in the LLA. Since the amplitudes in the LLA are real, only real parts of the NNLLA amplitudes are important. But even for these parts the pole Regge form becomes inapplicable because of the contributions of the  three-Reggeon cuts  which  appear in this approximation. Note that account of these  contributions also violates  unity  of  consideration of  quark-quark, quark-gluon and gluon-gluon scattering because the cuts give contributions to  amplitudes with different representations of the colour group in the $t$-channel for these processes. But even worse is that we actually do not know the contributions of the cuts.   

We showed  that the non-factorizing terms in the parton's elastic scattering amplitudes with the colour octet in the $t$-channel and negative signature calculated  using the infrared factorization  \cite{DelDuca:2014cya}  in two and three loops (non-logarithmic terms nonvanishing at $\epsilon\rightarrow 0$  in two loops and one-logarithmic terms singular at $\epsilon\rightarrow 0$ in three loops) can be explained  by the contribution of the three-Reggeon cut.  But these terms may  have other explanations. In particular, in \cite{Caron-Huot:2017fxr} for their explanation, along with the three-Reggeon cut,   mixing of the pole and the cut is used.  In these paper,  the coupling of the cut with partons and the Reggeon-cut mixing  were introduced using effective theory of Wilson lines.     
 
If we compare our current understanding of the cuts with the history of  investigation of the  gluon Reggeization, it seems that we are even in the worse position than after \cite{Lipatov:1976zz}, because the Regge form of elastic amplitudes was confirmed there  by direct calculation in two loops with power accuracy. Remind that to prove this form in all orders of perturbation theory it was firstly generalized for multiple production amplitudes in the multi-Regge kinematics, then the bootstrap conditions  for elastic and inelastic amplitudes  were derived,   after which it was proved that their fulfilment is sufficient for  justification of   the form,  and finally it was  shown that they are fulfilled.

\section{Conclusions}
The gluon Reggeization is one of the remarkable properties  of QCD. The Reggeization means  existence of the Regge pole with the gluon quantum numbers  and negative signature with the trajectory $j(t) = 1+\omega(t)$  with $\omega(0)=1$. It is important that this pole gives the main contribution to  QCD amplitudes with octet representation of the colour group in cross channels and negative signature. Thereby these amplitudes have a simple factorized form (pole Regge form)
in the leading   and  next-to-leading  logarithmic approximations (LLA and NLLA). This is true not only for elastic amplitudes, but also for real parts of multiparticle production  amplitudes  in the multi-Regge-kinematics.  

Regge cuts, generated by the Regge poles, appear in the amplitudes with the colour octet  and negative signature in the next-to-next-to-leading logarithmic 
approximation (NNLLA).  Strictly speaking, one can not assert that these amplitudes are completely given by contributions of the pole and the cut. A direct proof of such assertion could be coincidence of these contributions with results of  calculation of the amplitudes in the perturbation theory. Here we showed that for elastic scattering amplitudes  the coincidence  exists  for the non-logarithmic terms nonvanishing at $\epsilon\rightarrow 0$  in two loops and the one-logarithmic terms singular at $\epsilon\rightarrow 0$ in three loops.  

Contributions of the Regge cuts greatly complicates derivation of  the BFKL equation  in the NNLLA.  There are another  complication: the need to consider in the unitarity relations  products of real parts of amplitudes with negative signature in the NLLA, products of imaginary parts of  amplitudes with negative signature in the LLA, and products of amplitudes  with positive signature in the LLA. But they do not seem as significant as the need to take into account Regge cuts. 

\section{Acknowledgements}
V.S.F. thanks the Dipartimento di Fisica
dell'Universit\`{a} della Calabria and the Istituto Nazionale di
Fisica Nucleare (INFN), Gruppo Collegato di Cosenza, for warm hospitality
while part of this work was done and for financial support.

\end{document}